\documentclass[lettersize,journal]{IEEEtran}
\usepackage{amsmath,amsfonts}
\usepackage{algorithmic}
\usepackage{array}
\usepackage[caption=false,font=normalsize,labelfont=sf,textfont=sf]{subfig}
\usepackage{textcomp}
\usepackage{stfloats}
\usepackage{bm}
\usepackage{url}
\usepackage{verbatim}
\usepackage{algorithm}
\usepackage{booktabs}
\usepackage{graphicx}
\def\BibTeX{{\rm B\kern-.05em{\sc i\kern-.025em b}\kern-.08em
    T\kern-.1667em\lower.7ex\hbox{E}\kern-.125emX}}
\usepackage{balance}
\begin{document}

\title{Adaptive Residual Modeling for Scenario Learning in Probabilistic Load Forecasting}
\author{Jinbao Wang, Jun Liu, Shiliang Zhang, \IEEEmembership{Member, IEEE}, Xuehui Ma, Yushuai Li, \IEEEmembership{Senior Member, IEEE}, Lin Yan\thanks{J. Wang, J. Liu, X. Ma and  L. Yan (jbwang@stu.xaut.edu.cn, liujun0310@xaut.edu.cn, xuehui@xaut.edu.cn, yanlin@stu.xaut.edu.cn) are with Xi'an University of Technology, China. S. Zhang (shilianz@ifi.uio.no) is with University of Oslo, Norway. Y. Li (yusli@cs.aau.dk) is with Aalborg University, Denmark.} }

\maketitle

\begin{abstract}
Uncertainties in weather, user behavior, and distributed energy production introduce significant variability into feeder power. Effective power system operations require both multi-horizon load forecasts and well-calibrated extreme quantiles for the forecast. Nevertheless, standard objectives that fit the full forecast distribution often underemphasize rare observations and extremes. Conversely, applying heavy tail-weighting in the distribution to capture these extremes can distort the underlying probability distribution. To address this issue, this paper proposes adaptive residual modeling for scenario learning (ARMSL), a two-stage framework. Particularly, Stage I in our framework sequentially models environmental response, history-dependent load patterns, persistent residual deviations, and the remaining innovations. Upon the modeling, we then generate joint active- and reactive-power scenarios with state-dependent dispersion. Stage II keeps the scenario model fixed and calibrates extreme marginal quantiles using a simple, channel-specific reference forecast (an anchor). Only outward deviations from this anchor are admitted, with their contribution determined by local calibration history and the current operating state. From our numerical simulations on FeederBW dataset, Stage I outperforms diffusion baselines in energy and variogram scores. Our complete framework achieves an active-power 0.99-quantile pinball loss of 0.1932 (a 30.08\% to 85.33\% reduction relative to three baselines) with 1.054\% exceedance rate against a nominal 1\%. From simulations on the Rolle dataset, we observe decreased losses by 5.39\%--14.40\% compared to the active-power-only anchor. Furthermore, the unchanged Stage I scenarios in the simulation yield a correlation of 0.953 between predicted and observed feeder event rates, with 90.0\% of feeder errors falling within two percentage points.
\end{abstract}


\section{Introduction}
\label{sec:introduction}

\IEEEPARstart{W}{eather} dynamics and energy user behavior introduce significant uncertainty into photovoltaic (PV) output, electric vehicle (EV) charging, energy storage dispatch, and flexible demand. This variability propagates directly to aggregated active power $P$ and reactive power $Q$ measured at power distributions feeder. Consequently, operational decisions regarding reserves, demand-response, and voltage control must account for this uncertainty prior to observing future trajectories~\cite{terren2025extreme}. Different grid applications require different representations of this uncertainty. While marginal quantiles evaluate localized risk at individual forecast horizons, joint scenarios preserve the essential correlations across both time and power channels ($P$ and $Q$). These probabilistic forecasts provide essential input for downstream grid analyses to map uncertain power injections~\cite{wang2026hgnnppf}.

Probabilistic load forecasting typically employs quantile or Bayesian models to estimate marginal uncertainty, alongside scenario generators to simulate coherent joint trajectories~\cite{wang2024bayesiantransformer,zhao2025conditionaldiffusion}. However, three key challenges complicate extreme-quantile estimation in aggregate $P/Q$ forecasting, three challenges are particularly relevant to extreme quantiles: (i) a single feeder meter aggregates diverse physical processes operating across distinct temporal scales, (ii) loss functions fitted to the full forecast distribution tend to underemphasize rare tail errors, and (iii) the scarcity of extreme events at individual sites and horizons renders site-specific location calibration inherently unstable. Below we elaborate those issues and existing solutions.

The first challenge concerns the disentanglement of aggregate power variation. A feeder load trace reflects a combination of smooth environmental response, load history effects, persistent local deviations, and random dispersion. To isolate these dynamics, existing studies develop targeted feature-engineering or sample-weighting methods. \textit{E.g.}, Frequency-based reconstruction approach supports component-specific feature learning~\cite{wang2026solar}, while sample reweighting and model ensembling mitigate fitting biases and variance~\cite{wang2023reweighting,su2024robustensemble}. Similarly, drift-aware predictors adapt to non-stationary distributions, and physics-informed models enforce structural constrain~\cite{cao2024conceptdrift,liu2024physicsinformed}. While these techniques demonstrate the value of modular modeling, they do not prescribe a unified framework for stage-by-stage residual decomposition. Without an explicit sequence, unmodeled environmental dynamics can leak into lag components, and persistent deviations can inflate the estimated innovation variance. 

The second challenge arises from the inherent tension between joint-distribution modeling and extreme-quantile calibration. Standard Bayesian and shared-quantile predictors estimate marginal uncertainty across the full sample record~\cite{wang2024bayesiantransformer,ryu2024quantilemixer}, whereas conditional diffusion and dependence models generate coherent multi-horizon trajectories~\cite{zhao2025conditionaldiffusion,stover2024dependence}. Although weather-informed inputs and extreme-event generators enrich tail representation~\cite{zhang2025weatherinformed,wang2025extremegan}, abundant central observations continue to dominate objectives averaged across channels and horizons. Increasing the weight on tail errors can improve extreme quantile accuracy. However, forcing the joint model to fit extreme tails can warp its baseline representations and corrupt the spatial and temporal dependencies between $P$ and $Q$. 

The third challenge stems from the scarcity and temporal instability of extreme local observations. Since extreme events are inherently rare at a single feeder and horizon, fitting localized tail parameters directly often yields high varance and unstable quantile bounds. Existing research attempts to stablize quantile estimation through several strategies: Sharing parameters across probability levels~\cite{ryu2024quantilemixer,luo2024qrnn}, pooling complementary models via forecast combinations~\cite{yang2024cplfs,cao2025combination}, and enforcing noncrossing constraints to maintain logical quantile order~\cite{chen2025noncrossing,huang2024noncrossing}. To handle non-stationary dynamics, peak-density and online methods dynamically adjust to distribution shifts~\cite{he2023peakdensity,qin2025shift}. However, these approaches primarily rely on global pooling or recent temporal trends alone. Thus, they fail to evaluate whether historical tail discrepancies remain physically relevant under the feeder's real-time operating state.

To address these three challenges, we propose adaptive residual modeling for scenario learning (ARMSL), a unified two-stage framework. Stage I addresses aggregated variation (the first challenge) by sequentially isolating environmental response, history-dependent memory, persistent residual shocks (inferred deviations), and joint innovation distributions, recombining them into complete multi-horizon $P/Q$ scenarios. Stage II decouples extreme calibration from scenario generation (the second challenge) by holding the Stage I model fixed. To overcome local tail scarcity and dynamic relevance (the third challenge), Stage II introduces a channel-specific \textit{anchor} forecast to supply candidate discrepancies from the scenario model. We also design a \textit{retention coefficient} to dictates the baseline proportion of outward discrepancies to admit based on calibration history, while real-time operating state features dynamically modulate their relevance. To maintain physical and statistical validity, these state-adptive corrections are strictly governed by directional, adjacent-quantile, and cumulative ordering constraints that guarantee non-crossing quantile bounds. We summarize the main contributions of our work as follows.

\begin{enumerate}
 \item We propose a structured scenario model that isolates load components in a prescribed temporal order. Environmental dynamics are extracted prior to fitting a regularized lag kernel for load history dependence. Remaining variations are captured via group-sparse shocks and asymmetric, state-scaled innovations to generate coherent multi-channel trajectories.
 \item We design a two-stage architecture that strictly separate joint-distribution estimation from extreme-quantile calibration. Since Stage II cannot modify Stage I parameters or scenarios, extreme-tail adjustments are guaranteed to leave the central probability mass and cross-variable dependence structure intact.
 \item We introduce a tail calibration mechanism that merges historical calibration evidence with real-time operating state. Site-level retention with regional fallback governs baseline anchor contributions, while bounded state mappings adjust dynamics relevance. The directional constraints exclude inward anchor adjustments, and adjacent-quantile limits with cumulative ordering enforce a valid, non-crossing quantile grid.
\end{enumerate}

In the remaining of this paper, we elaborate our two-stage framework in Section~\ref{sec:method}, and provide detailed report on numerical simulations in assessing our design in Section~\ref{sec:experiments}. We conclude this paper in Section~\ref{sec:conclusion}.

\section{Methodology}
\label{sec:method}

This section details the two-stage frame work for the proposed adaptive residual modeling for scenario learning (ARMSL). In ARMSL, Stage I estimates the joint conditional distribution of future $P/Q$ trajectories. Stage II calibrates their extreme marginal quantiles using information from channel-specific anchors, local history, and the current state. The Stage I scenario generator remains fixed throughout Stage II. Below we brief the load forecast problem to be solved by ARMSL.

Let $s\in\mathcal S$ denote a feeder or aggregate measurement site, $t$ the forecast issuance time, and $\Delta$ the sampling interval. The inputs are observed active power $P$, reactive power $Q$, temperature $T$, solar irradiance $R$, and a calendar vector $\boldsymbol\tau$ containing time-of-day, day-of-week, seasonal, weekend, and holiday features. Define $\boldsymbol Y_{s,t}=[P_{s,t},Q_{s,t}]^{\mathsf T}$ and $\boldsymbol U_{s,t}=[T_{s,t},R_{s,t}]^{\mathsf T}$. All predictors are measurable with respect to $\mathcal H_{s,t}$, the information available by time $t$.

For the ordered horizon set $\mathcal H_f=\{h_1,\ldots,h_H\}$ with dimension $H$, let $\boldsymbol y=[\boldsymbol y_1^{\mathsf T},\ldots,\boldsymbol y_H^{\mathsf T}]^{\mathsf T}$ denote a possible future trajectory. The target is the $2H$-dimensional conditional distribution
\begin{equation}
 F_{s,t}(\bm y)=\Pr\!\left(\bm Y_{s,t+h_1}\leq\bm y_1,\ldots,
 \bm Y_{s,t+h_H}\leq\bm y_H\mid\mathcal H_{s,t}\right),
 \label{eq:target}
\end{equation}
where the inequalities are componentwise and $H=|\mathcal H_f|$. Both channels and all horizons are forecast directly, without recursively feeding predictions into subsequent forecasts. ARMSL uses the decomposition
\begin{equation}
 \bm Y_{s,t+h}=\bm{\mathcal E}_{s,t,h}+\bm{\mathcal M}_{s,t,h}
 +\bm J_{s,t,h}+\bm\epsilon_{s,t,h}.
 \label{eq:decomposition}
\end{equation}
The components represent environmental and calendar response ($\boldsymbol E$), distributed-lag memory ($\boldsymbol M$), persistent local residual shocks ($\boldsymbol J$), and the remaining correlated innovation ($\boldsymbol\epsilon$). Each fitted component is held fixed while its residual is passed to the next component, except for the single memory refit described below. Smoothness, lag decay, group sparsity, and heavy-tailed dispersion distinguish their statistical roles.

In the proposed framework, the recombination produces the Stage I scenario core, meaning the fitted joint generator and its sampled trajectories. Stage II takes the core's empirical lower and upper extreme quantiles for $P$ and $Q$ as calibration bases. Below we elaborate the two stages in detail.

\subsection{Stage I: Joint Scenario Modeling in Residual Order}

Three residual handoffs organize Stage I. Particluarly, the environmental fitting removes repeatable weather and calendar effects. The distributed-lag model then extracts dependence on the residual history, and the shock model captures persistent local deviations. The remaining residuals determine the joint innovation distribution.

\subsubsection{Environmental response}

Site-specific month-hour medians, estimated from the training period, define reference temperature and irradiance. Their differences from current observations are the anomalies $\Delta T_{s,t}$ and $\Delta R_{s,t}$. Four irradiance phases---night, sunrise, daytime, and sunset---distinguish rising and receding solar conditions at similar irradiance. Continuous inputs are standardized using training data only.

Predetermined causal filters summarize past environmental conditions. For $v\in\{T,R\}$ and smoothing scale $\lambda\in\Lambda_v$,
\begin{equation}
 Z^{(\lambda)}_{v,s,t}=\frac{1}{C_{v,\lambda}}
 \sum_{\ell=0}^{L_e}\exp(-\ell\Delta/\lambda)\,\Delta v_{s,t-\ell},
 \label{eq:weather_state}
\end{equation}
where $L_e$ is the maximum environmental lag, $C_{v,\lambda}$ normalizes the weights, and $\Lambda_v$ is the set of candidate smoothing scales. Current observations, anomalies, differences over multiple scales, irradiance phase, and filtered summaries form the environmental feature vector $\boldsymbol z^U_{s,t}$.

A low-capacity additive model estimates the repeatable conditional location:
\begin{equation}
\begin{aligned}
 \bm{\mathcal E}_{s,t,h}={}&\bm b_{s,h}(\bm\tau_t)
 +\sum_j\bm f_{h,j}(z^U_{s,t,j})\\
 &+\sum_{(j,k)\in\mathcal I}\bm f_{h,jk}(z^U_{s,t,j},z^U_{s,t,k}).
\end{aligned}
 \label{eq:environment}
\end{equation}
In (\ref{eq:environment}), $\boldsymbol b_{s,h}$ contains low-order daily, weekly, and annual Fourier terms, together with weekend and holiday indicators. Indices $j$ and $k$ select environmental features. Each $\boldsymbol f_{h,j}$ is linear or a low-degree spline, and $\mathcal I$ restricts interactions to temperature-time, irradiance-time, temperature-season, and temperature-irradiance pairs. Ridge and spline-curvature penalties limit flexibility so that the environmental model does not absorb rapid variation intended for later components. The two power channels share features but have separate coefficients. The environmental residual is $\boldsymbol R^E_{s,t}=\boldsymbol Y_{s,t}-\boldsymbol E_{s,t,0}$.

\subsubsection{History-dependent residual memory over physical lag}

The memory model uses more than the latest environmental residual. The same current residual can follow a sustained rise, a plateau, or recovery from a peak, with different implications for the next trajectory. Thermal inertia, production continuity, charging duration, and control recovery motivate such history dependence at aggregate meters. This is the sense in which the load memory is termed non-Markov: It is not restricted to a first-order model of the latest residual. Cubic B-splines $B_b$, $b=1,\ldots,B$, summarize the history on a logarithmic physical-lag axis, up to maximum lag $L_y$:
\begin{equation}
\begin{aligned}
 \bm m_{s,b,t}&=\sum_{\ell=1}^{L_y}B_b\!\left(\log[1+\ell\Delta]\right)
 \bm R^E_{s,t-\ell},\\
 \bm{\mathcal M}_{s,t,h}&=\sum_{b=1}^{B}\bm A_{s,h,b}(\bm z^U_{s,t})\bm m_{s,b,t}.
\end{aligned}
 \label{eq:memory}
\end{equation}
Each $2\times2$ matrix $\boldsymbol A_{s,h,b}$ maps the two-channel history to a forecast contribution. Diagonal entries describe within-channel effects, and off-diagonal entries describe predictive cross-channel associations. Environmental modulation is enabled only for lag scales accepted by time-ordered validation. It changes response strength while keeping the lag basis and physical-lag axis fixed.

For continuous physical lag $\xi$, the local kernel is $\boldsymbol K_{s,h}(\xi)=\boldsymbol K_h^{\mathrm{shared}}(\xi)+\delta\boldsymbol K_{s,h}(\xi)$. Regional sufficient statistics estimate the shared kernel, and stronger shrinkage is applied to the site-specific deviation. Stacking the site-, horizon-, and basis-specific matrices into $\boldsymbol A$ gives the regularizer
\begin{equation}
\begin{aligned}
 \Omega_K={}&\lambda_{K,2}\|\bm D^2\bm A\|_F^2
 +\lambda_{K,d}\sum_b w_b\|\bm A_b\|_F^2\\
 &+\lambda_{K,g}\sum_b\|\bm A_b\|_F
 +\lambda_{K,\times}\sum_b\|\operatorname{offdiag}(\bm A_b)\|_F^2,
\end{aligned}
 \label{eq:memory_penalty}
\end{equation}
where $\bm D^2$ is the second-difference operator along logarithmic lag, $\boldsymbol A_b$ is the coefficient slice for basis $b$, and $w_b$ increases with physical lag. The operator $\operatorname{offdiag}(\cdot)$ retains only off-diagonal entries. The four terms penalize roughness, distant-lag effects, unnecessary basis groups, and cross-channel coefficients, respectively. These penalties favor gradual history effects. Subtracting memory yields $\boldsymbol R^M_{s,t,h}=\boldsymbol Y_{s,t+h}-\boldsymbol E_{s,t,h}-\boldsymbol M_{s,t,h}$.

\subsubsection{Persistent residual shocks}

The residual $\boldsymbol R^M_{s,t,h}$ contains both abrupt local structure and nonpersistent noise. Indexing by target time $u=t+h$ aligns residuals associated with the same observation. A componentwise median over at least two valid horizons gives the two-channel consensus residual $\boldsymbol r^{\mathrm{cons}}_{s,u}$. For $N_{\mathrm{tr}}$ training times, the shock sequence is estimated by
\begin{equation}
\begin{aligned}
\widehat{\bm J}_{s,1:N_{\mathrm{tr}}}=\arg\min_{\bm J}&\frac12\sum_u
 \|\bm r^{\mathrm{cons}}_{s,u}-\bm J_{s,u}\|_2^2
 +\lambda_{J,g}\sum_u\|\bm J_{s,u}\|_2\\
 &+\lambda_{J,\mathrm{TV}}\sum_u
 \|\bm J_{s,u}-\bm J_{s,u-1}\|_1.
\end{aligned}
 \label{eq:shock}
\end{equation}
The group penalty $\lambda_{J,g}$ promotes sparse activation of the two-channel shock vector. The total-variation penalty $\lambda_{J,\mathrm{TV}}$ favors persistence of its amplitude over neighboring times. Together, they represent short persistent clusters separately from the distributed-lag memory. Online estimation uses a one-step proximal update initialized from the previous causal state. Later observations therefore do not revise previously issued shock estimates.

After shock extraction, one damped refit updates the memory coefficients using nonshock training samples, with the basis and penalties fixed. Let $S_{s,t}\in\{0,1\}$ denote the causal shock state. Future activation and active-state amplitude follow
\begin{equation}
\begin{aligned}
 \Pr(S_{s,t+\Delta}=1\mid S_{s,t}=0)&=p_{\mathrm{on}},\\
 \Pr(S_{s,t+\Delta}=1\mid S_{s,t}=1)&=p_{\mathrm{cont}},\\
 \bm J_{s,t+\Delta}&=\rho_{J,s}\bm J_{s,t}+\bm\eta_{s,t+\Delta}.
\end{aligned}
 \label{eq:shock_transition}
\end{equation}
In (\ref{eq:shock_transition}), $p_{\mathrm{on}}$ and $p_{\mathrm{cont}}$ are activation and continuation probabilities, $\rho_{J,s}$ controls site-specific persistence, and $\boldsymbol\eta$ is the amplitude innovation in an active state. Inactive states set $\boldsymbol J$ to zero. The shock dynamics are Markov, while the memory term in (\ref{eq:memory}) retains the longer load history. Subtraction gives $\boldsymbol R^J_{s,t,h}=\boldsymbol R^M_{s,t,h}-\widehat{\boldsymbol J}_{s,t,h}$, the residual used to estimate the continuous innovation distribution.

\subsubsection{Joint innovations and scenario reconstruction}

Stacking $\boldsymbol R^J$ across both channels and all horizons gives a $D=2H$ innovation vector. Index $d\in\{1,\ldots,D\}$ identifies a channel-horizon pair. Draw $\boldsymbol z\sim t_\nu(\boldsymbol0,\boldsymbol R)$ from a centered $D$-variate Student-$t$ distribution, where $\nu$ is the degrees of freedom and $\boldsymbol R\in\mathbb R^{D\times D}$ is a shrinkage estimate of the correlation parameter. The initial innovation is
\begin{equation}
 \epsilon^{(0)}_{s,t,d}=\mu_{s,d}+
 \begin{cases}
  \sigma^-_{s,d}z_d, & z_d<0,\\
  \sigma^+_{s,d}z_d, & z_d\geq0,
 \end{cases}
 \label{eq:innovation}
\end{equation}
where location $\mu_{s,d}$ is the training residual median. Robust lower and upper scales $\sigma^-_{s,d}$ and $\sigma^+_{s,d}$ are estimated from quantile differences. Their asymmetry permits different dispersion in the two directions. Student-$t$ tails accommodate isolated large errors, while $\boldsymbol R$ represents dependence across channels and horizons.

Recent operating conditions modify dispersion through
\begin{equation}
\begin{aligned}
 c_{s,t,d}&=\exp\!\left\{\operatorname{clip}
 (\widetilde{\bm v}_{s,t}^{\mathsf T}\bm\beta_{s,d},a_{\min},a_{\max})\right\},\\
 \epsilon_{s,t,d}&=\mu_{s,d}+c_{s,t,d}
 (\epsilon^{(0)}_{s,t,d}-\mu_{s,d}).
\end{aligned}
 \label{eq:dynamic_scale}
\end{equation}
The standardized state $\widetilde{\boldsymbol v}_{s,t}$ contains daily and annual cyclic features, absolute weather anomalies, and causal $P/Q$ volatility summaries over multiple scales. The function $\operatorname{clip}(x,a,b)$ truncates $x$ to $[a,b]$. Regularized coefficients $\boldsymbol\beta_{s,d}$ link this state to dimension $d$, and $a_{\min}$ and $a_{\max}$ bound the logarithmic scale multiplier. The fitted multiplier expands dispersion in states associated with larger training innovations and contracts it in steadier states. When causal volatility history is unavailable, $c_{s,t,d}=1$.

For scenario $m\in\{1,\ldots,M\}$, the components are recombined as
\begin{equation}
 \bm Y^{(m)}_{s,t+h}=\bm{\mathcal E}_{s,t,h}+\bm{\mathcal M}_{s,t,h}
 +\bm J^{(m)}_{s,t,h}+\bm\epsilon^{(m)}_{s,t,h}.
 \label{eq:scenario}
\end{equation}
Each draw spans both power channels and all horizons. Let $q^C_{\alpha,s,t,h,c}$ be its empirical marginal quantile at probability level $\alpha$ for channel $c\in\mathcal C=\{P,Q\}$. Increasing the scenario count $M$ reduces Monte Carlo quantile-estimation error without changing the fitted generator. The trajectories support decisions that depend on joint behavior. In Stage II, we will use their extreme marginal quantiles as fixed calibration inputs.

\subsection{Stage II: State-Constrained Tail Calibration}

Rare observations provide limited information about the core's extreme marginal quantiles. Fitting these endpoints separately can produce variable estimates or crossing quantiles, whereas retraining the generator with a stronger tail objective can change its joint distribution. Stage II therefore keeps the core fixed and combines simple channel-specific anchors with historical retention and bounded state-dependent corrections. The directional constraint applies to the anchor contribution, and the separate core-bias correction can move either inward or outward. Below we explain the procedures in Stage II in detail.

Let $\mathcal A=\{\alpha_1<\cdots<\alpha_K\}$, $K\geq3$, be the obtained quantile grid, and let $\mathcal T=\{\tau_L,\tau_U\}$ contain its extremes, with $\tau_L=\alpha_1$ and $\tau_U=\alpha_K$. Case index $i$ identifies a site, issuance time, and horizon. The signs $d_{\tau_L}=-1$ and $d_{\tau_U}=1$ define outward movement. An outward anchor contribution has the form $d_\tau\Delta_{\tau,i,c}$ with $\Delta_{\tau,i,c}\geq0$. Interior levels retain the core quantiles.

\subsubsection{Candidate tail evidence and hierarchical retention}

An anchor is a low-capacity reference forecast for the same power channel, using only information available at issuance. For active power, ridge autoregression uses the history vector $\boldsymbol p^A_{s,t}=[P_{s,t},P_{s,t-\Delta},\ldots,P_{s,t-L_A}]^{\mathsf T}$:
\begin{equation}
 \widehat P^A_{s,t,h}=b^A_{s,h}+(\bm\beta^A_{s,h})^{\mathsf T}
 \operatorname{Std}_{\mathrm{tr}}(\bm p^A_{s,t}),
 \label{eq:p_anchor}
\end{equation}
where $L_A$ is the history length, $b^A_{s,h}$ is the intercept, $\boldsymbol\beta^A_{s,h}$ contains ridge coefficients, and $\operatorname{Std}_{\mathrm{tr}}$ applies stored training standardization. The reactive-power location is the persistence forecast $\widehat Q^A_{s,t,h}=Q_{s,t}$. For either point anchor $\widehat Y^A_{s,t,h,c}$, define the calibration residual $e^A_{s,t,h,c}=Y_{s,t+h,c}-\widehat Y^A_{s,t,h,c}$. Adding its empirical $\tau$-quantile to the point forecast gives $q^A_{\tau,s,t,h,c}$.

The candidate evidence is the anchor's outward discrepancy from the core:
\begin{equation}
 G_{\tau,i,c}=\left[d_\tau(q^A_{\tau,i,c}-q^C_{\tau,i,c})\right]_+,
 \qquad [x]_+=\max(x,0).
 \label{eq:anchor_gap}
\end{equation}
If the lower-tail anchor is above the core's lower endpoint, or the upper-tail anchor is below its upper endpoint, the corresponding gap is zero. Thus, anchor evidence can expand a tail but cannot narrow it.

Calibration pinball loss selects a base retention coefficient from $\Lambda_\lambda\subset[0,1]$:
\begin{equation}
 \widehat\lambda^{\mathrm{site}}_{\tau,s,h,c}=\arg\min_{\lambda\in\Lambda_\lambda}
 \frac{1}{|\mathcal I_{s,h}|}\sum_{i\in\mathcal I_{s,h}}
 \rho_\tau(y_{i,c},q^C_{\tau,i,c}+d_\tau\lambda G_{\tau,i,c}),
 \label{eq:retention}
\end{equation}
The set $\mathcal I_{s,h}$ contains valid calibration cases for site $s$ and horizon $h$. The pinball loss is $\rho_\tau(y,q)=(\tau-\mathbb I\{y<q\})(y-q)$, where $\mathbb I$ is the indicator function. A site uses its fitted coefficient when at least $N_{\min}$ cases are available. Otherwise, it uses a regional coefficient for the same horizon, channel, and tail. Denote the resulting hierarchical coefficient by $\lambda^{(0)}_{\tau,s,h,c}$. The base endpoint is
\begin{equation}
 q^B_{\tau,i,c}=q^C_{\tau,i,c}+d_\tau
 \lambda^{(0)}_{\tau,s_i,h_i,c}G_{\tau,i,c}.
 \label{eq:base_tail}
\end{equation}
Regional fallback limits the variability of estimates based on sparse local data. This base retention summarizes historical calibration evidence, and its relevance to an individual forecast is then adjusted using the current state.

\subsubsection{State-conditioned evidence admission}

The issuance state includes current power, changes over a predetermined lag $\delta_r$, and daily and weekly cyclic features:
\begin{equation}
\begin{aligned}
 \bm x_{s,t}=[&P_{s,t},Q_{s,t},P_{s,t}-P_{s,t-\delta_r},
 Q_{s,t}-Q_{s,t-\delta_r},\\
 &\sin\omega_t,\cos\omega_t,\sin\upsilon_t,\cos\upsilon_t]^{\mathsf T},
\end{aligned}
 \label{eq:tail_state}
\end{equation}
The variables $\omega_t$ and $\upsilon_t$ encode within-day and within-week phases. Recent changes distinguish rising and receding trajectories with similar current power. For case $i$, $\boldsymbol x_i$ denotes the standardized state.

Let $a_{\tau_L}=\alpha_2$ and $a_{\tau_U}=\alpha_{K-1}$ denote the adjacent interior probability levels. The local tail scale is
\begin{equation}
 S_{\tau,i,c}=\max\!\left\{d_\tau(q^C_{\tau,i,c}-q^C_{a_\tau,i,c}),
 G_{\tau,i,c},\varepsilon_s(|Y_{s_i,t_i,c}|+1)\right\},
 \label{eq:local_scale}
\end{equation}
where $\varepsilon_s>0$ supplies a site-specific numerical floor. Three zero-initialized linear maps adjust the core endpoint, anchor-gap magnitude, and retention:
\begin{equation}
\begin{aligned}
 q^{J,C}_{\tau,i,c}=d_\tau\max\!\big\{&d_\tau q^C_{a_\tau,i,c},\\
 &d_\tau[q^C_{\tau,i,c}+c_J\tanh(g_{C,\tau,c}(\bm x_i))
 S_{\tau,i,c}]\big\},\\
 G^J_{\tau,i,c}&=G_{\tau,i,c}\exp\{c_J\tanh(g_{A,\tau,c}(\bm x_i))\},\\
 \lambda^J_{\tau,i,c}=\mathrm{clip}\big[&\lambda^{(0)}_{\tau,s_i,h_i,c}\\[-0.15em]
 &+c_J\tanh(g_{\lambda,\tau,c}(\bm x_i)),0,1\big].
\end{aligned}
 \label{eq:branches}
\end{equation}
The validated capacity $c_J>0$ bounds the normalized core shift, logarithmic gap multiplier, and retention increment before clipping. Each channel and tail has its own maps $g_C$, $g_A$, and $g_\lambda$. Superscript $J$ marks Stage II adjustments and is distinct from the Stage I shock variable $\boldsymbol J$. The first branch limits the corrected endpoint relative to the adjacent interior quantile. It can move inward or outward within that limit. The second rescales a nonnegative outward gap, and the third keeps the retained fraction within $[0,1]$.

Combining the three branches gives
\begin{equation}
 \widehat q^{\mathrm{final}}_{\tau,i,c}=q^{J,C}_{\tau,i,c}
 +d_\tau\lambda^J_{\tau,i,c}G^J_{\tau,i,c}.
 \label{eq:final_endpoint}
\end{equation}

The anchor term in (\ref{eq:final_endpoint}) is always outward and vanishes when $G_{\tau,i,c}=0$. Small retention keeps the output close to the state-adjusted core endpoint. Note that this directional guarantee concerns the anchor term, rather than the net displacement from the original core quantile.

\subsubsection{Tail objective and ordered quantile output}

Interior levels retain $\widehat q^{\mathrm{final}}_{\alpha,i,c}=q^C_{\alpha,i,c}$ for $\alpha\in\mathcal A\setminus\mathcal T$. The empirical core grid is ordered, and the adjacent-level limits in (\ref{eq:branches}) keep the corrected tails outside the interior grid. A cumulative maximum provides an ordering safeguard:
\begin{equation}
 \widetilde q^{\mathrm{final}}_{\alpha_j,i,c}=\max_{1\leq k\leq j}
 \widehat q^{\mathrm{final}}_{\alpha_k,i,c},\qquad j=1,\ldots,K.
 \label{eq:projection}
\end{equation}
The four channel-tail endpoints are fitted jointly by minimizing
\begin{equation}
\begin{aligned}
 \mathcal L_{\mathrm{tail}}={}&\frac{1}{|\mathcal C||\mathcal T|N}
 \sum_{c\in\mathcal C}\sum_{\tau\in\mathcal T}\sum_{i=1}^{N}
 \frac{\rho_\tau(y_{i,c},\widetilde q^{\mathrm{final}}_{\tau,i,c})}{\kappa_c}\\
 &+\lambda_J^{\mathrm{tail}}\sum_{\bm p\in\Theta_J}
 \frac{\|\bm p\|_2^2}{|\bm p|},
\end{aligned}
 \label{eq:tail_loss}
\end{equation}
where $N$ is the number of valid calibration cases, $y_{i,c}$ is the observed target, $\kappa_c$ is a robust scale of $|Y_c|$, and $\Theta_J$ contains the trainable branch parameters. The symbol $|\boldsymbol p|$ denotes the number of elements in parameter array $\boldsymbol p$, and $\lambda_J^{\mathrm{tail}}$ is the weight-decay coefficient. Channel normalization prevents the larger power scale from dominating the objective. The endpoints share state features but use separate channel- and tail-specific parameters. Directional gaps, bounded maps, adjacent-level limits, and cumulative ordering constrain their outputs.

Algorithm~\ref{alg:armsl_pipeline} summarizes estimation and forecasting. Training $\mathcal D_{\mathrm{tr}}$, validation $\mathcal D_{\mathrm{val}}$, and tail calibration $\mathcal D_{\mathrm{cal}}$ use nonoverlapping partitions. Stage I is fitted and frozen before Stage II calibration. At issuance, only $\mathcal H_{s,t}$ is used. The calibrated quantiles are separate outputs from the unchanged joint scenarios.

\begin{algorithm}[H]
\caption{ARMSL estimation and forecast generation}
\label{alg:armsl_pipeline}
\footnotesize
\begin{algorithmic}[1]
\REQUIRE Data sets $\mathcal D_{\mathrm{tr}}$, $\mathcal D_{\mathrm{val}}$, $\mathcal D_{\mathrm{cal}}$, history $\mathcal H_{s,t}$, horizons $\mathcal H_f$, and scenario count $M$
\ENSURE Joint scenarios $\{\bm Y_{s,t+h}^{(m)}\}$ and ordered quantiles $\{\widehat q^{\mathrm{final}}_{\alpha,s,t,h,c}\}$
\STATE \textbf{Stage I -- fit the residual-ordered scenario core}
\STATE Fit the environmental response $\bm{\mathcal E}$ on the causally prepared $\mathcal D_{\mathrm{tr}}$
\STATE Form $\bm R^E$ and estimate distributed-lag memory $\bm{\mathcal M}$
\STATE Extract shocks $\bm J$ and refit $\bm{\mathcal M}$ once on nonshock samples
\STATE Fit shock dynamics, innovation margins, correlation $\bm R$, and dynamic scale
\STATE Select penalties and distributional parameters on $\mathcal D_{\mathrm{val}}$
\STATE Freeze the Stage I scenario core
\STATE \textbf{Stage II -- fit the state-constrained tail calibrator}
\STATE Fit anchors, hierarchical retention, and bounded maps on $\mathcal D_{\mathrm{cal}}$
\FOR{each issuance $(s,t)$}
 \STATE \textbf{Stage I -- generate joint scenarios}
 \STATE Apply the stored causal transforms and lag kernels to $\mathcal H_{s,t}$
 \STATE Update the current shock state and simulate $M$ future shock paths
 \STATE Draw correlated dynamic innovations and reconstruct $M$ scenarios by (\ref{eq:scenario})
 \STATE \textbf{Stage II -- calibrate extreme endpoints}
 \STATE Extract scenario-core quantiles and evaluate the Stage II endpoint maps
 \STATE Project the reported quantile grid to monotone order by (\ref{eq:projection})
\ENDFOR
\RETURN $\{\bm Y_{s,t+h}^{(m)}\}$ and $\{\widehat q^{\mathrm{final}}_{\alpha,s,t,h,c}\}$
\end{algorithmic}
\end{algorithm}

\section{Experimental Evaluation and Numerical Results}
\label{sec:experiments}

\subsection{Experimental Setup}

\textit{Platform and data.} Our simulations uses FeederBW dataset, which contains two years of one-minute measurements from 200 low-voltage feeders in Germany, hourly weather fields, and feeder metadata \cite{treutlein2026real}. Active and reactive power are averaged to 10-min resolution. Direct forecasts are issued at 10, 20, 30, and 60 min. The common training cutoff is March 31, 2024, at 22:00 UTC, followed by 42,508 contiguous calibration cases and 567,908 evaluation cases.

We also use the Rolle dataset, which contains power measurements and numerical weather predictions for 24 anonymized meters in a Swiss distribution grid\cite{nespoli2019rolle}. The experiments use 10-min records from January 13, 2018, to January 19, 2019. Data before 2018-09-01 is used to  train Stage I, and the next two months provide validation. Within the remaining test period, the first 28 local days calibrate Stage II and the following 53-day data is used to evaluate it.

In the simulations, all series are aligned to a common time grid, and power is averaged over completed intervals. Environmental averages become available only after their source intervals end. Splits are defined by target time $t+h$. Imputation uses no future observations, and cases without the required causal history are excluded. Means, scales, spline knots, and month-hour climate references are estimated from training data only. Sparse climate cells fall back to the training median for the same hour, then to the global training median. Both datasets use the same variables, model structure, and evaluation protocol, with all transforms and parameters estimated separately for each dataset.

\textit{Baselines and common protocol.} ARMSL is compared with the weather-informed Temporal Fusion Transformer (WI-TFT)~\cite{zhang2025weatherinformed}, the conditional diffusion model denoted DiffLoad in these experiments~\cite{zhao2025conditionaldiffusion}, and the optical-flow denoising diffusion Transformer (OFDDT)~\cite{yao2025ofddt}. Each architecture is adapted to direct multihorizon $P/Q$ forecasting. Stage II calibrates $\mathcal T=\{0.01,0.99\}$ for both channels. The neural baselines use PyTorch 2.4.1 and CUDA 12.1. Stage II runs on CPU with the scenario core fixed. All methods use the same issuance times, targets, causal inputs, and evaluation cases. Site-balanced summaries give sites equal weight. Joint scores use 128 scenarios per issuance, and endpoint estimates use 8,192 samples.

\textit{Metrics and statistical inference.} Let $N$ be the number of evaluation cases, $\boldsymbol y_i\in\mathbb R^D$ the observed trajectory, and $\boldsymbol y_i^{(m)}$ its $m$th scenario. Marginal quality is measured by the continuous ranked probability score (CRPS), $\mathrm{CRPS}_{i,d}=\int_{-\infty}^{\infty}[\widehat F_{i,d}(z)-\mathbb I\{y_{i,d}\leq z\}]^2\,\mathrm dz$, where $\widehat F_{i,d}$ is the empirical marginal cumulative distribution. Joint quality is measured by the energy score and the variogram score with $p=0.5$:
\begin{equation}
\begin{aligned}
 \mathrm{ES}_i={}&\frac1M\sum_m\|\bm y_i^{(m)}-\bm y_i\|_2
 -\frac{1}{2M^2}\sum_{m,r}\|\bm y_i^{(m)}-\bm y_i^{(r)}\|_2,\\
 \mathrm{VS}_i={}&\sum_{a<b}w_{ab}\left(|y_{i,a}-y_{i,b}|^p
 -\frac1M\sum_m|y_{i,a}^{(m)}-y_{i,b}^{(m)}|^p\right)^2,
\end{aligned}
\label{eq:exp_joint_metrics}
\end{equation}
where $m$ and $r$ index scenarios, $a$ and $b$ index trajectory dimensions, $w_{ab}$ is a fixed nonnegative pair weight, and $\|\cdot\|_2$ is the Euclidean norm. Endpoint loss, upper-tail exceedance, relative loss improvement, and event-probability error are defined as
\begin{equation}
\begin{aligned}
 \rho_\tau(y,q)&=(\tau-\mathbb I\{y<q\})(y-q),\\
 \widehat e_\tau&=N^{-1}\sum_i\mathbb I\{y_i>q_{\tau,i}\},\\
 G(m,b)&=100(1-L_m/L_b)\%,\\
 \mathrm{BS}&=N^{-1}\sum_i(\widehat\pi_i-o_i)^2,
\end{aligned}
\label{eq:exp_tail_metrics}
\end{equation}
where $q_{\tau,i}$ is the forecast quantile, and $L_m$ and $L_b$ are site-balanced losses for method $m$ and baseline $b$ on the same evaluation cases. The Brier score (BS) compares predicted event probability $\widehat\pi_i$ with binary outcome $o_i$. Lower losses are preferred, and $\widehat e_{0.99}$ should approach 1\%. Interval coverage and normalized width describe calibration and sharpness separately. Operational tests also report the miss rate at a fixed alert budget and reserve at a fixed violation rate.

Confidence intervals (CIs) use 5,000 site-clustered percentile bootstrap replicates. Simulations on Rolle dataset additionally uses date-clustered and two-way site-date checks. Operational comparisons use 10,000 feeder-clustered replicates, paired Wilcoxon tests, and Holm correction. Percentage changes and confidence limits are reported to two decimal places, and case proportions to one decimal place. 

\subsection{Integrated Two-Stage Forecasting Performance}
\label{sec:results}

\begin{figure}[!t]
 \centering
 \includegraphics[width=0.78\columnwidth]{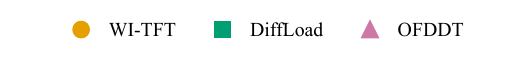}\\[-0.45em]
 \begin{minipage}[t]{0.485\columnwidth}
  \vspace{0pt}\centering
  \includegraphics[width=\linewidth]{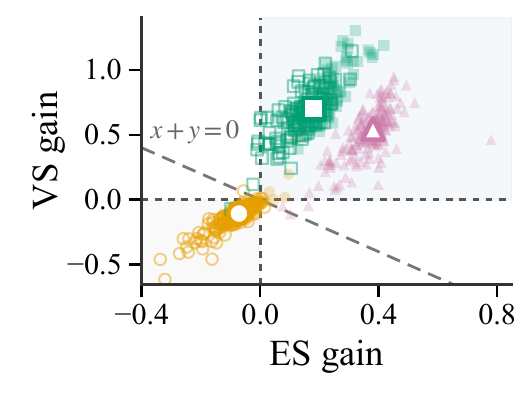}\\[-0.25em]
  {\footnotesize (a) Stage I: joint distribution}
 \end{minipage}\hfill
 \begin{minipage}[t]{0.485\columnwidth}
  \vspace{0pt}\centering
  \includegraphics[width=\linewidth]{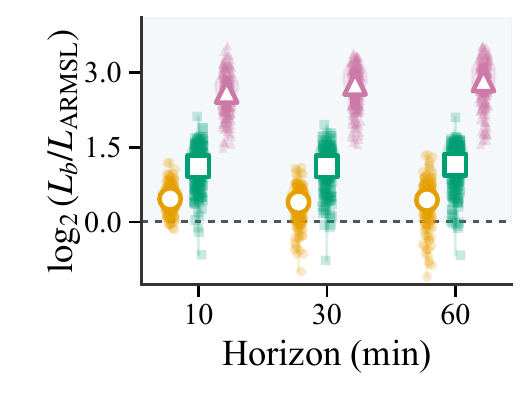}\\[-0.25em]
  {\footnotesize (b) Stage II: active-power tail}
 \end{minipage}
 \caption{Two-stage evaluation on FeederBW (200 feeders). (a) Joint-score gains; filled points also indicate CRPS improvement. (b) Active-power 0.99-quantile loss ratios. Positive log-ratios favor ARMSL. Large markers denote medians; bars in (b) show interquartile ranges.}
 \label{fig:integrated_performance}
\end{figure}

Figure~\ref{fig:integrated_performance} shows the evaluation results for the two stages on the same test cases. Stage I provides joint trajectories and core quantiles, and Stage II calibrates the four extreme marginal endpoints without changing the scenarios. The diagnostic $\log_2(L_b/L_a)$ compares baseline loss $L_b$ with ARMSL loss $L_a$, so positive values favor ARMSL.

The results show that Stage I reduces energy and variogram scores by 11.02\% and 37.46\% relative to DiffLoad, and by 23.58\% and 28.49\% relative to OFDDT. Mean CRPS also decreases by 23.51\% in the diffusion-baseline comparison. Both joint scores improve for 97.0\% of feeders against DiffLoad and 98.5\% against OFDDT (Fig.~\ref{fig:integrated_performance}(a)). Positive combined log-score gains across the two comparisons occur for 99.5\% of feeders. We also observe that WI-TFT achieves lower scores for the full distribution. ARMSL's principal advantage over this baseline is at the extreme endpoints, and that the larger relative variogram improvement over DiffLoad is consistent with improved representation of differences across channels and horizons.

The complete framework achieves an active-power 0.99-quantile pinball loss of 0.1932 (Table~\ref{tab:tail}), which is 30.08\%, 54.27\%, and 85.33\% below WI-TFT, DiffLoad, and OFDDT, respectively. The reduction relative to WI-TFT has a site-clustered 95\% CI of 27.40\%--32.50\%, and horizon-specific reductions range from 27.13\% to 33.52\%. Figure~\ref{fig:integrated_performance}(b) shows the endpoint comparison. Across paired $P/Q$ cases, ARMSL has lower loss in 88.6\%, 97.7\%, and 100.0\% of cases against the three baselines, respectively. All reported group medians and interquartile ranges favor ARMSL. Reactive-power reductions are 21.60\%--28.40\% relative to WI-TFT and 48.60\%--53.30\% relative to DiffLoad.

These paired gains are consistent with state-dependent use of channel-specific anchor evidence. Current power, recent changes, and cyclic features regulate the correction, while the channel-normalized objective balances $P$ and $Q$. If no outward gap exists, the anchor contributes zero and the endpoint equals the state-adjusted core value.

\begin{table}[!t]
 \centering
 \caption{Feeder-balanced FeederBW results for the active-power 0.99 quantile. Nominal exceedance is 1\%; reserve and shortage are normalized metrics.}
 \label{tab:tail}
 \setlength{\tabcolsep}{3.4pt}
 \begin{tabular}{lrrrr}
  \toprule
  Method & Pinball & Exceed./\% & Reserve & Shortage\\
  \midrule
  ARMSL & \textbf{0.1932} & 1.054 & 1.060 & 0.00344\\
  WI-TFT & 0.2764 & 6.946 & 0.463 & 0.01515\\
  DiffLoad & 0.4225 & 0.126 & 2.939 & 0.00037\\
  OFDDT & 1.3170 & 36.897 & 0.076 & 0.09065\\
  \bottomrule
 \end{tabular}
\end{table}

\begin{figure}[!t]
 \centering
 \includegraphics[width=0.9\columnwidth]{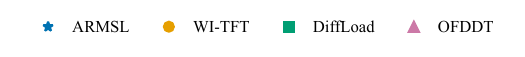}\\[-0.45em]
 \includegraphics[width=0.9\columnwidth]{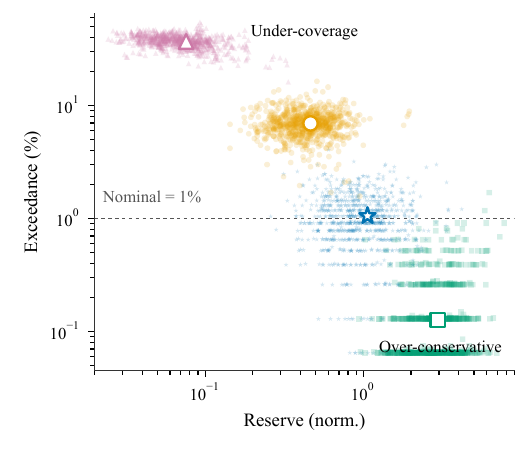}
 \caption{Active-power 0.99-quantile exceedance versus normalized reserve on FeederBW (800 feeder-horizon cases per method). Large markers summarize each method. The dashed line marks nominal 1\% exceedance. Zero counts are displayed at half-event resolution on the logarithmic axis.}
 \label{fig:calibration_sharpness}
\end{figure}

We observe that the ARMSL exceedance rate of 1.054\% is close to the nominal 1\%. WI-TFT and OFDDT provide insufficient reserve in this comparison, whereas DiffLoad uses 2.77 times the ARMSL reserve and has much lower exceedance. ARMSL reduces normalized shortage by 77.29\% relative to WI-TFT and normalized reserve by 63.93\% relative to DiffLoad. Figure~\ref{fig:calibration_sharpness} shows this calibration-reserve tradeoff across 800 feeder-horizon cases. These reductions compare different exceedance outcomes, and the matched-risk comparison is reported separately in Section~III-E.

\begin{figure}[!t]
 \centering
 \includegraphics[width=\columnwidth]{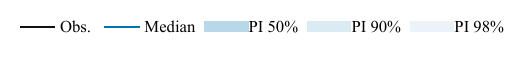}\\[-0.45em]
 \begin{minipage}{0.485\columnwidth}\centering
  \includegraphics[width=\linewidth]{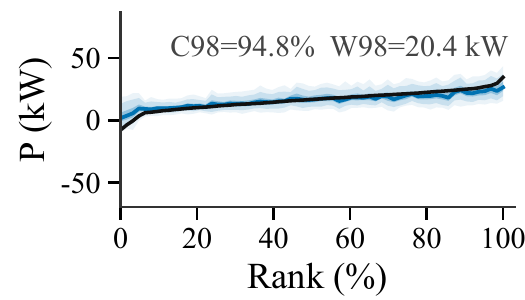}\\[-0.25em]
  {\footnotesize (a) ARMSL}
 \end{minipage}\hfill
 \begin{minipage}{0.485\columnwidth}\centering
  \includegraphics[width=\linewidth]{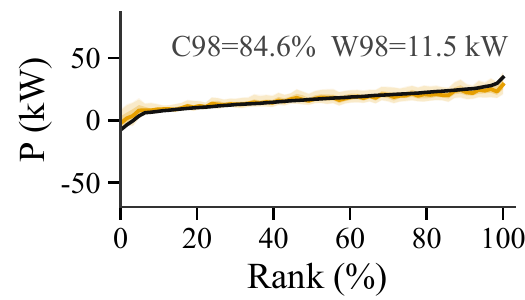}\\[-0.25em]
  {\footnotesize (b) WI-TFT}
 \end{minipage}\\[-0.1em]
 \begin{minipage}{0.485\columnwidth}\centering
  \includegraphics[width=\linewidth]{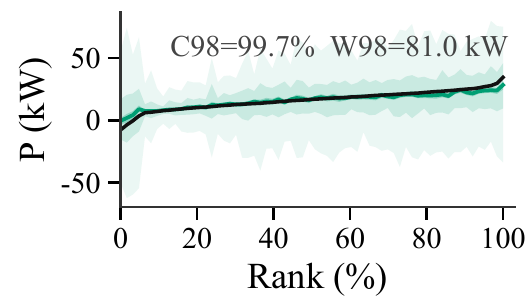}\\[-0.25em]
  {\footnotesize (c) DiffLoad}
 \end{minipage}\hfill
 \begin{minipage}{0.485\columnwidth}\centering
  \includegraphics[width=\linewidth]{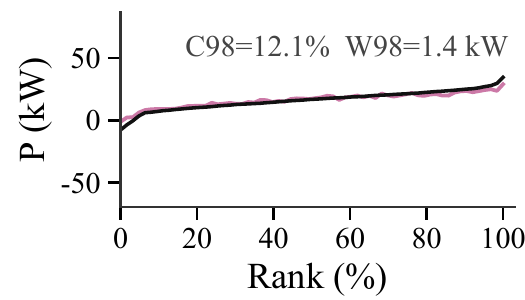}\\[-0.25em]
  {\footnotesize (d) OFDDT}
 \end{minipage}
 \caption{Risk-ordered 98\% prediction intervals at 60 min for a representative feeder. Panels use the same vertical limits. $C_{98}$ and $W_{98}$ denote empirical coverage and mean interval width.}
 \label{fig:risk_intervals}
\end{figure}

The results show that directional anchor admission, bounded corrections, and quantile-order constraints limit the correction according to the issuance state. Figure~\ref{fig:risk_intervals} illustrates 60-min, 98\% prediction intervals for the feeder with the median ARMSL gain over WI-TFT among the 200 feeders. ARMSL attains 94.8\% coverage with a mean width of 20.4~kW. WI-TFT gives 84.6\% coverage with a narrower interval, while DiffLoad gives 99.7\% at 81.0~kW, and OFDDT covers 12.1\%. Thus, the ARMSL interval is closer to nominal coverage than WI-TFT's but still undercovers in this example. The dynamic innovation scale adapts spread to recent volatility before Stage II calibration.

\begin{figure}[!t]
 \centering
 \includegraphics[width=0.9\columnwidth]{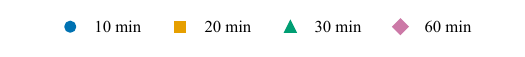}\\[-0.45em]
 \includegraphics[width=0.9\columnwidth]{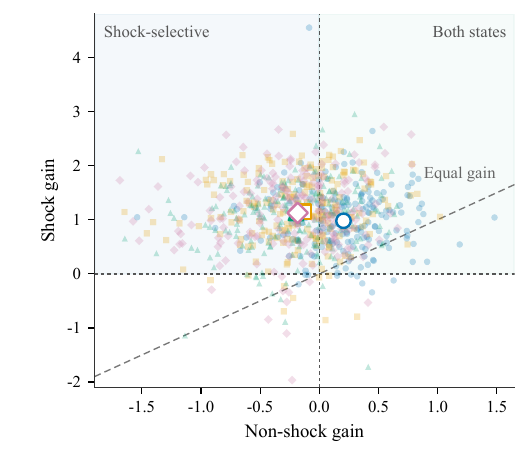}
 \caption{Tail-loss gains over WI-TFT in inferred shock and nonshock states (200 feeders per horizon). Large markers show horizon medians. Points above the diagonal have larger gains during shocks.}
 \label{fig:shock_gain}
\end{figure}

ARMSL and WI-TFT losses are paired by feeder, horizon, and inferred shock state on the same evaluation cases. During shocks, reductions range from 49.40\% to 55.70\% over the 10--60-min horizons, compared with $-15.60\%$ to 12.50\% outside shocks. In Figure~\ref{fig:shock_gain}, 94.8\% of feeder-horizon pairs lie above the equal-gain diagonal, and 53.0\% improve during shocks while deteriorating outside them. This pattern is consistent with larger benefits when Stage I propagates persistent deviations and Stage II responds to current power and recent changes. 

\subsection{Evaluation on a Second Regional Dataset}

Simulations on Rolle dataset evaluate the calibration mechanism on a different population, year, and load distribution. The ARMSL structure is retained, but all parameters are re-estimated locally, including 28 days of causal tail calibration. Note that it tests performance after local fitting rather than transfer of fixed parameters. At 10, 20, 30, and 60 min, loss reductions relative to the active-power-only anchor are 7.61\%, 7.25\%, 8.72\%, and 11.31\% on FeederBW, and 5.39\%, 5.83\%, 6.29\%, and 14.40\% on Rolle. Site-clustered CIs have positive lower bounds.

\begin{figure}[!t]
 \centering
 \includegraphics[width=\columnwidth]{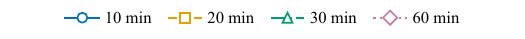}\\[-0.45em]
 \begin{minipage}[t]{0.485\columnwidth}
  \vspace{0pt}\centering
  \includegraphics[width=\linewidth]{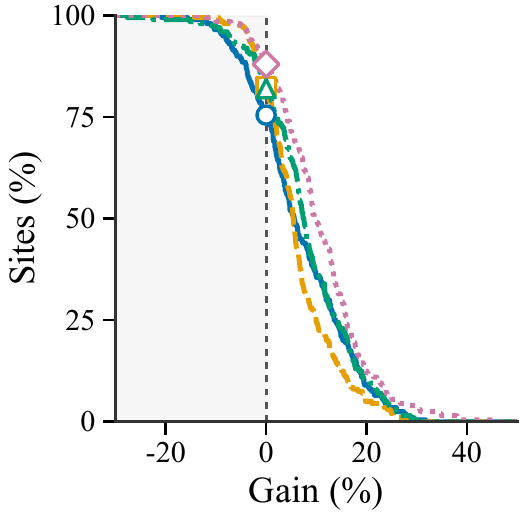}\\[-0.25em]
  {\footnotesize (a) FeederBW}
 \end{minipage}\hfill
 \begin{minipage}[t]{0.485\columnwidth}
  \vspace{0pt}\centering
  \includegraphics[width=\linewidth]{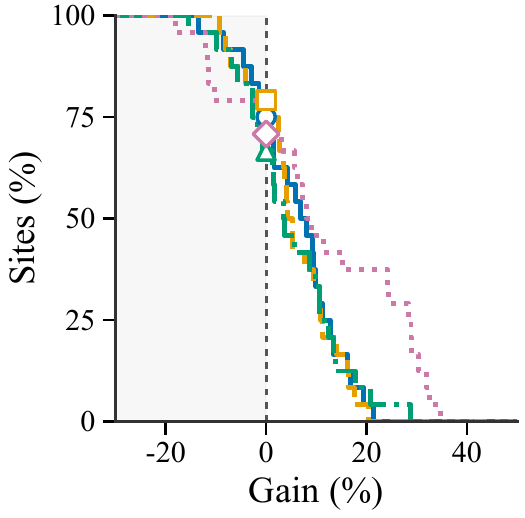}\\[-0.25em]
  {\footnotesize (b) Rolle}
 \end{minipage}
 \caption{Fraction of sites exceeding each gain threshold relative to P-only on (a) FeederBW (200 sites) and (b) Rolle (24 sites). Symbols at zero indicate win rates. Rolle losses are averaged over three seeds.}
 \label{fig:site_heterogeneity}
\end{figure}

Figure~\ref{fig:site_heterogeneity} shows the fraction of sites whose gain exceeds each threshold, where $g_s=100(1-L_{\mathrm{ARMSL},s}/L_{P\text{-only},s})$. At zero, the curves give win rates of 75.5\%--88.0\% on FeederBW and 66.7\%--79.2\% on Rolle. All four horizons improve for 120 of 200 FeederBW feeders and 14 of 24 Rolle sites. The three Rolle seeds agree on the direction of the reported gains. Local histories determine the anchors and retention estimates, with regional fallback for sites having fewer calibration cases.

\subsection{Mechanism Ablations}

The simulations show that replacing the state multiplier with a static scale increases energy and variogram scores by 2.07\% and 10.61\% on FeederBW, and by 1.52\% and 7.42\% on Rolle. During inferred shocks, dynamic scaling improves these scores by 3.36\% and 14.55\% on FeederBW, and by 2.54\% and 10.18\% on Rolle. State-dependent dispersion can accommodate volatility changes that a fixed scale cannot. The larger relative variogram response is consistent with improved variation across channels and horizons.

The following ablations use the same evaluation cases. \emph{Core} denotes the Stage I quantiles, \emph{P-only} denotes direct use of the active-power-only anchor, and \emph{Dynamic} denotes the dynamic-retention variant evaluated before the final constraints. Relative to Core, P-only gives gains of 1.33\%, 0.15\%, $-1.95\%$, and $-5.08\%$ at the four horizons. Dynamic gives 8.81\%, 7.36\%, 6.92\%, and 6.80\%, reversing the losses at 30 and 60 min.

\begin{figure}[!t]
 \centering
 \includegraphics[width=\columnwidth]{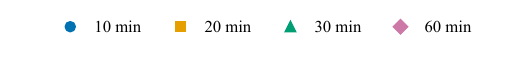}\\[-0.45em]
 \begin{minipage}[t]{0.485\columnwidth}
  \vspace{0pt}\centering
  \includegraphics[width=\linewidth]{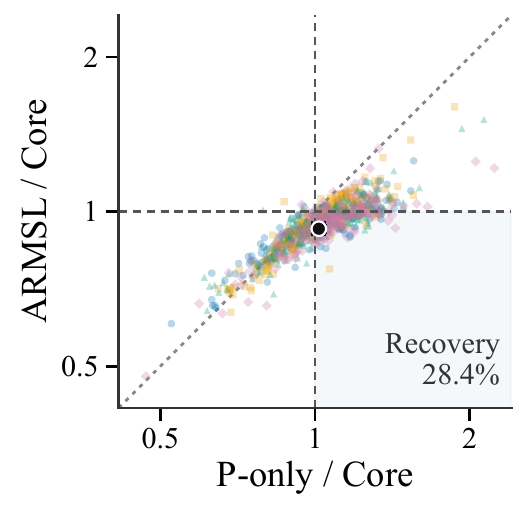}\\[-0.25em]
  {\footnotesize (a) Loss comparison}
 \end{minipage}\hfill
 \begin{minipage}[t]{0.485\columnwidth}
  \vspace{0pt}\centering
  \includegraphics[width=\linewidth]{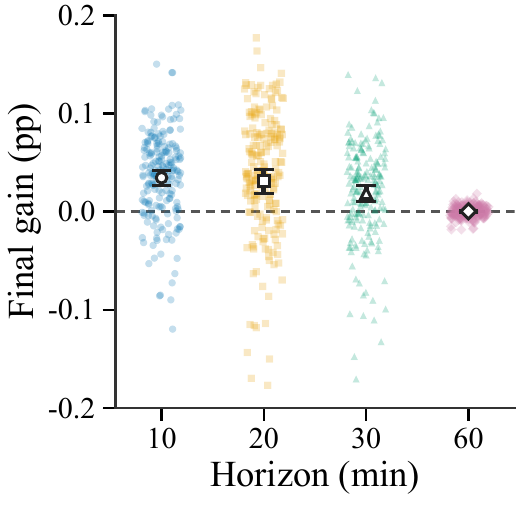}\\[-0.25em]
  {\footnotesize (b) Final step}
 \end{minipage}
 \caption{Component effects on FeederBW (200 feeders per horizon). (a) Paired loss ratios; the black marker summarizes all horizons. (b) Increment from the final constraints. Large markers and bars show mean-loss summaries and 95\% paired-bootstrap CIs.}
 \label{fig:component_ablation}
\end{figure}

Figure~\ref{fig:component_ablation}(a) pairs P-only/Core and ARMSL/Core loss ratios on logarithmic axes. Points below the diagonal favor ARMSL. The recovery quadrant contains 28.4\% of feeder-horizon pairs: Direct anchor substitution worsens Core, but ARMSL improves it. The black marker summarizes equal-feeder, equal-horizon mean losses using 5,000 paired feeder-cluster bootstrap replicates. Panel~(b) reports the final increment, $100(L_{\mathrm{Dynamic},s}-L_{\mathrm{ARMSL},s})/L_{\mathrm{Core},s}$, and summaries divide mean loss differences by mean Core loss.

Direct substitution discards the core's environmental, memory, shock, and joint-innovation information. Dynamic retention instead retains the core and incorporates an outward anchor discrepancy according to the current state. The final constraints preserve the direction of the anchor gap and quantile order. Their additional mean gain on FeederBW is less than 0.04 percentage points.

Table~\ref{tab:anchor_uncertainty} summarizes 72 seed-specific gain intervals on Rolle: 36 for the unconstrained variant and 36 for ARMSL. Resampling uses sites (S), dates (D), or both ($\mathrm S\times\mathrm D$). Unconstrained mean gains over P-only are 5.09\%, 3.21\%, 8.06\%, and 14.98\% at the four horizons. Positive lower bounds occur in 36/36 ARMSL intervals and 15/36 unconstrained intervals. A positive $L_{\min}$ means that all three seed-specific lower bounds are positive for that horizon and resampling scheme. ARMSL reduces mean interval width under all three schemes, despite smaller point gains at 30 and 60 min. The seeds share 24 sites and 53 dates, so these width averages are descriptive summaries of separate intervals, not pooled CIs. Together with the small final increment on FeederBW, the results support empirical stability on Rolle.

\begin{table}[!t]
 \centering
 \caption{Rolle uncertainty across three seeds. $L_{\min}$ is the smallest lower endpoint of the seed-specific 95\% CIs; $\overline W$ is their mean width. Intervals are summarized separately, not pooled.}
 \label{tab:anchor_uncertainty}
 \footnotesize
\setlength{\tabcolsep}{2.8pt}
\renewcommand{\arraystretch}{1.03}
\begin{tabular*}{\columnwidth}{@{\extracolsep{\fill}}rlrrrr@{}}
\toprule
$h$ (min) & & \multicolumn{2}{c}{Unconstrained} & \multicolumn{2}{c}{ARMSL} \\
\cmidrule(lr){3-4}\cmidrule(lr){5-6}
& Unit & $L_{\min}$ (\%) & $\bar W$ (pp) & $L_{\min}$ (\%) & $\bar W$ (pp) \\
\midrule
10 & S & -2.42 & 13.86 & 1.55 & 7.61 \\
 & D & 0.19 & 9.35 & 2.65 & 5.34 \\
 & $\mathrm{S}\times\mathrm{D}$ & -5.07 & 18.28 & 0.15 & 10.41 \\
\addlinespace[2pt]
20 & S & -3.77 & 12.69 & 2.05 & 7.35 \\
 & D & -1.50 & 9.32 & 2.51 & 6.51 \\
 & $\mathrm{S}\times\mathrm{D}$ & -6.52 & 17.84 & 0.38 & 11.22 \\
\addlinespace[2pt]
30 & S & -1.03 & 16.59 & 2.14 & 8.26 \\
 & D & 4.10 & 7.73 & 2.62 & 7.27 \\
 & $\mathrm{S}\times\mathrm{D}$ & -2.76 & 19.73 & 0.21 & 12.49 \\
\addlinespace[2pt]
60 & S & 3.85 & 20.89 & 6.58 & 14.40 \\
 & D & 9.97 & 9.40 & 10.80 & 6.70 \\
 & $\mathrm{S}\times\mathrm{D}$ & 2.33 & 24.26 & 5.61 & 16.64 \\
\bottomrule
\end{tabular*}
\end{table}

\subsection{Operational Decision Value}

\begin{table*}[tbh]
 \centering
 \caption{Computational resources on FeederBW. ARMSL measures incremental Stage II calibration; baselines measure complete fitting. GPU memory is the increase above prelaunch idle.}
 \label{tab:resource_cost}
 \scriptsize
 \setlength{\tabcolsep}{4.2pt}
 \begin{tabular}{lrrrrrrr}
  \toprule
  Method & ACU (\%) & AMC (MB) & AGMU (MB) & AGUR (\%) & TTD (min) & AET (s) & SIT (ms) \\
  \midrule
  ARMSL Stage II & 3.30 & 1774.97 & 0.00 & 0.00 & 2.37 & 0.002 & 0.032 \\
  DiffLoad & 3.12 & 3548.75 & 782.70 & 40.49 & 94.66 & 74.72 & 0.433 \\
  OFDDT & 3.03 & 3610.76 & 1372.06 & 71.79 & 69.29 & 94.09 & 30.249 \\
  WI-TFT & 3.10 & 3529.25 & 1530.15 & 55.90 & 214.64 & 128.61 & 3.512 \\
  \bottomrule
 \end{tabular}
\end{table*}

Operational evaluation uses the trajectory maximum of apparent power, $S^{\max}_{s,t}=\max_{h\in\mathcal H_f}\sqrt{P_{s,t+h}^2+Q_{s,t+h}^2}$. Feeder-specific training quantiles at 0.95 and 0.99 define event thresholds. Each feeder contributes at most 768 issuances, and 128 scenarios estimate event probabilities and capacity quantiles. These tests use the unchanged Stage I scenarios, and Stage II endpoint corrections do not enter the sampled trajectory maximum. Dependence ablations retain matched marginals while removing cross-horizon dependence or all multivariate dependence. A separate P-only ablation converts active-power scenarios to apparent power using the issuance-time power factor.

\begin{figure}[!t]
 \centering
 \includegraphics[width=0.9\columnwidth]{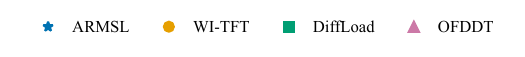}\\[-0.5em]
 \includegraphics[width=0.9\columnwidth]{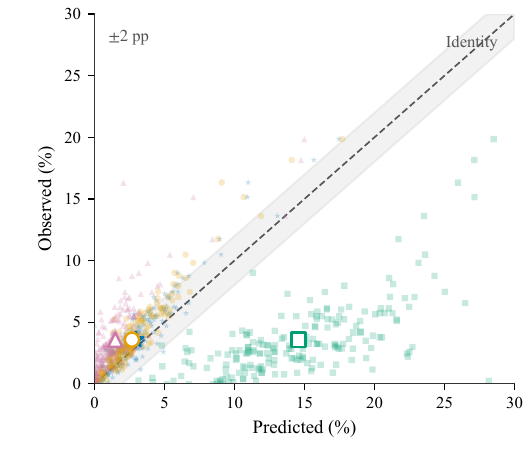}
 \caption{Predicted versus observed high-load event rates at the training 0.99 threshold (200 feeders per method). Large markers show means. The gray band extends two percentage points on either side of the identity line.}
 \label{fig:event_probability}
\end{figure}

At the 0.99 threshold, the observed event rate for the four-horizon maximum is 3.58\%, compared with the 1\% exceedance associated with an individual marginal threshold. ARMSL predicts 2.95\%. Figure~\ref{fig:event_probability} shows a correlation of 0.953 between predicted and observed feeder event rates, with 90.0\% of errors within two percentage points. The corresponding proportions are 86.0\% for WI-TFT, 0.5\% for DiffLoad, and 53.5\% for OFDDT. DiffLoad predicts 14.56\%, while OFDDT predicts 1.48\%. Broad marginal tails can increase the maximum-event probability, while dependence determines how exceedances coincide across horizons. 

Feeder reserve is the issuance-mean positive capacity margin above the point forecast, normalized by the feeder's training event threshold. For each feeder and model, reserve is linearly interpolated to empirical violation rates of 3\%, 5\%, and 7\%, using service levels of 0.90, 0.95, 0.975, and 0.99. Comparisons retain paired feeders whose curves span the target rate, and no extrapolation is used. This is a retrospective comparison at matched empirical risk.

\begin{figure}[!t]
 \centering
 \includegraphics[width=0.9\columnwidth]{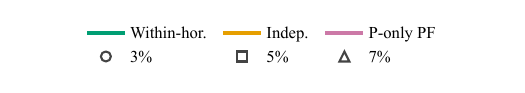}\\[-0.45em]
 \includegraphics[width=0.9\columnwidth]{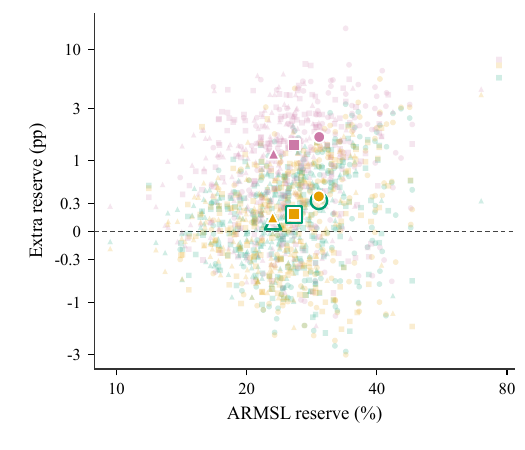}
 \caption{Reserve at matched empirical risk for the training 0.99 threshold (148--187 paired feeders per comparison). Small markers denote individual feeders; large markers denote paired means. Positive extra reserve indicates lower reserve for ARMSL.}
 \label{fig:reserve_value}
\end{figure}

At the training 0.99 threshold, ARMSL saves 0.330, 0.172, and 0.104 percentage points of threshold-normalized reserve relative to the within-horizon $P/Q$ ablation at the three violation targets. Savings are 0.386, 0.181, and 0.145 percentage points relative to independent marginals, and 1.663, 1.410, and 1.160 relative to fixed-power-factor P-only scenarios. All paired feeder-cluster CIs for mean savings are positive, although individual feeders can have negative savings. Figure~\ref{fig:reserve_value} retains these negative effects. Its axes show $100R_{\mathrm{ARMSL},s}$ and $100(R_{b,s}-R_{\mathrm{ARMSL},s})$, using logarithmic and inverse-hyperbolic-sine spacing with ticks in the original units.

Within-horizon $P/Q$ dependence affects apparent-power tails. Cross-horizon dependence affects whether high values occur together or create separate opportunities for the maximum to exceed a threshold. The fixed-power-factor ablation omits future reactive-power sampling. The reserve results are consistent with the value of retaining these relationships in joint shock paths and correlated innovations. At a 20\% alert budget, ARMSL lowers the miss rate by 0.49 and 1.17 percentage points relative to the within-horizon and fixed-power-factor ablations, respectively.

\subsection{Computational Cost}

Table~\ref{tab:resource_cost} reports resource measurements from one workstation. The ARMSL row measures incremental Stage II calibration, whereas the baseline rows measure complete model fitting. Process trees are sampled every second, and inference time is measured per complete multihorizon issuance. Average CPU utilization (ACU) is the share of logical CPU capacity, and average memory consumption (AMC) is resident memory. Under the Windows Display Driver Model (WDDM), average GPU memory use (AGMU) is measured above prelaunch idle, and average GPU utilization rate (AGUR) measures device activity. The remaining columns are total training duration (TTD), average epoch time (AET), and single inference time (SIT). The results show Stage II fits in 2.37~min on the CPU, uses 1774.97~MB of memory, and takes 0.032~ms per multihorizon issuance. It optimizes 112 parameters using fixed core endpoints. These values quantify the additional cost of tail calibration. 

\section{Conclusion}
\label{sec:conclusion}

This paper developed ARMSL to combine joint multihorizon $P/Q$ forecasting with separate extreme-quantile calibration, in two stages. Stage I models environmental response, distributed-lag memory, and persistent residual shocks, then generates trajectories with correlated asymmetric innovations and state-dependent dispersion. Stage II keeps these scenarios fixed. Channel-specific anchors provide outward candidate gaps, site-specific retention with regional fallback summarizes historical evidence, and bounded state maps adjust core bias, gap magnitude, and retention. Adjacent-quantile limits and cumulative ordering constrain the reported quantiles.

The findings from our evaluations highlight the importance of evaluating joint forecast quality and extreme-quantile reliability. ARMSL's tail improvements occur during inferred shock states, suggesting that operating-state information helps identify when endpoint corrections are most useful. The ablation simulations support preserving the scenario core and selectively incorporating anchor evidence, with constrained corrections improving stability when local tail observations are scarce. Results on Rolle support this design after local re-estimation, extending the evidence to a different feeder population and load distribution. The operational comparisons further show that retaining dependence across power channels and horizons can reduce reserve requirements at matched empirical risk. Together, these findings support separating trajectory generation from marginal tail calibration: Joint scenarios retain the structure needed for operational decisions, while a calibration layer refines extreme risk estimates at a small incremental computational cost.

An immediate future work it to examine hierarchical transfer to shorten Rolle's 28-day local calibration and test forecasting across regions without re-estimating parameters. Extending endpoint calibration to complete trajectories will require a constrained transformation that preserves relevant $P/Q$ and cross-horizon dependence. We also envision further evaluation to compare inferred shock states with disturbance and device-operation records, and measure end-to-end training and inference costs.

\bibliographystyle{IEEEtran}
\bibliography{references}

\end{document}